# Topological Study of the $H_3^{++}$ Molecular System: $H_3^{++}$ as a Cornerstone for Building Molecules during the Big Bang

by

**Bijit Mukherjee,**[(1)] **Debasis Mukhopadhyay,**[(2)]
**Satrajit Adhikari,**[‡(1)] **and Michael Baer**[*(3)]

[(1)] **Department of Physical Chemistry, Indian Association for Cultivation of Science, Jadavpur, Kolkata – 700 032, India**

[(2)] **Department of Chemistry, University of Calcutta, Kolkata 700 009, India**

[(3)] **The Fritz Haber Center for Molecular Dynamics, The Hebrew University of Jerusalem, Jerusalem 91904, Israel**

*Email: michaelb@fh.huji.ac.il

‡Email: pcsa@iacs.res.in

# Abstract

The present study is devoted to the possibility that tri-atomic molecules were formed during or shortly after the Big Bang. For this purpose we consider the ordinary $H_3^+$ and $H_3$ molecular systems and the primitive tri-atomic molecular system, $H_3^{++}$, which, as is shown, behaves differently. The study is carried out by comparing the topological features of these systems as they are reflected through their non-adiabatic coupling terms. Although $H_3^{++}$ is not known to exist as a molecule, we found that it behaves as such at intermediate distances. However this illusion breaks down as its asymptotic region is reached. Our study indicates that whereas $H_3^+$ and $H_3$ dissociate smoothly, the $H_3^{++}$ does not seem to do so. Nevertheless, the fact that $H_3^{++}$ is capable of living as a molecule on *borrowed* time enables it to catch an electron and form a molecule via the reaction $H_3^{++} + e \rightarrow H_3^+$ that may dissociate properly:

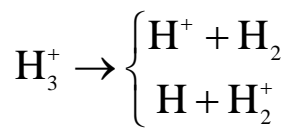

$$H_3^+ \rightarrow \begin{cases} H^+ + H_2 \\ H + H_2^+ \end{cases}$$

Thus, the two unique features acquired by $H_3^{++}$, namely, that it is the most primitive system formed by three protons and *one* electron and topologically, still remain for an instant a molecule, may make it the sole candidate for becoming the *cornerstone* for creating the molecules.

# I.   Introduction

This article is written under the influence of Professor Hawking's TED lecture, [1] presented several years ago, which discusses various issues associated with the Big Bang. In that lecture molecules were mentioned as eventually having been created *too early*. Following that we thought that such an idea could be supported by the quantum mechanical approach to studying molecules, as derived by the Born-Oppenheimer-Huang (BOH) theory.[2,3] This theory is known to be responsible for revealing the Non-Adiabatic Coupling Term (NACT) magnitudes found later to be, frequently, singular, [4,5] in fact poles, [6a] and therefore responsible for the quantization within molecules [6b,7a,8,20] reminiscent of the Bohr-Sommerfeld quantization in atoms. It is important to mention that two well-known effects are associated with the NACTS, namely the Jahn-Teller effect [6c,7,9-21] and the Renner-Teller effect. [22-24]

It is important to emphasize that NACTs can be created by any aggregate of *three* or more nuclei, immaterial if the aggregate is a stable molecule or just a weakly bound group of nuclei or even a group of protons moving inside an electronic cloud.

In this article we intend to study the possibility that molecules were, eventually, formed already at the earliest stages of the construction of the universe, namely, during or immediately after the Big Bang. It will be claimed that the necessary conditions for that to happen are the above mentioned NACTs. According the BOH approach molecules exist *if and only if* they are capable of switching from one adiabatic state to another and this process is, indeed, guaranteed by the NACTs. Thus instead of trying to find out how molecules are formed we have to justify, at least mathematically, the formation of the NACTs. This seems to be guaranteed since the BOH theory furnishes, mathematically, the *singular* NACTs [4,5,6a,7,8] and therefore, like numerous other particles in physics known to be characterized (mathematically) by singularities, it is expected to be formed during the Big Bang.

Still for that to happen we need a situation where *at least* three nuclei (e.g., protons) are at reasonable distances from each other, thus forming, e.g. a loosely bound ion, $H_3^{++}$ (lasting at least an instant of time) which, due to the electronic cloud, could lead at a somewhat later stage to the loosely bound $(H, H_2)^+$ ionic system which may end up with the formation of $H_2$, $H_2^+$ or $H_3^+$. Returning to the NACTs and their mathematical singularities that eventually hold together the loosely bound $H_3^{++}$ ion, if indeed this happens in the physical world, they are expected to form an origin (sink) for molecules.

Because of all that we intend to concentrate mainly on the $H_3^{++}$ system which may not be a real molecule but nevertheless, as will be shown, fulfills the BOH requirements and therefore can still be considered as a molecular system. Saying all that, studying the $H_3^{++}$ can be very intriguing because, indeed, it is the most 'primitive' tri-atomic/ionic system and as such can be considered as the '*mother*' of all tri-atomic/ionic molecular systems and eventually of all kinds of molecular systems, in general.

Part of our efforts will be devoted also to the much more popular and widely studied molecular systems, $H_3$ [6{e,f,g,h},25-29] and $H_3^+$ [30-41]. This we do because most of these studies were carried out while all three nuclei are rather close to each other – on the average about 3 a.u. apart (or less). In contrast, the present studies are extended up to 15 a.u. hoping to create extreme conditions for the existence of the NACTs.

## II. Theoretical Background

Our study relates to a planar system containing three particles (i.e., protons) defined in terms of three Jacobi coordinates ($r, R, \theta$) where $r$ is inter-atomic distance between two protons, $R$ is the distance between the third proton and the center-of-mass of the two previous protons and $\theta$ is the angle in-between them (in the present study it is assumed to be $\pi/2$). In addition we apply a second system of (polar) coordinates ($q, \varphi$) with its origin at some point on the $R$-axis – in most cases at the D$_{3h}$ point. Here $q$ is the radial coordinate and $\varphi$ is the corresponding angular coordinate (see Fig.1).

The numerical study centers on the three lower states of these molecular systems and consequently we refer to two kinds of conical intersections (*ci*): one kind formed by the two lower states and therefore all magnitudes related to it are designated as (1, 2) and the second kind formed by the two upper states, so that their magnitudes are designated as (2, 3). At this stage we remind the reader that the whole study is carried out for the planar configuration where the plane is formed by the three nuclei (for details see Section III).

The primary magnitudes to be studied in this article are the NACTs, always calculated along closed *planar* contours. It is well known that deriving accurate meaningful NACTs is a long and tedious task which requires repeated assessments and tests. One of the more crucial tests for their relevance is the fulfillment of the molecular *quantization* rule. Although this feature of the NACTs is frequently pointed out here it is mentioned again because of its importance for the present publication. Occasionally we distinguish between two-state results and tri-state results but most of the final results are derived for the tri-state case (most of the details related to the tri-state system are given in Appendix II).

In case of two, quasi-isolated, states formed by two adjacent states, namely, $j$ and $j+1$, the *quantization* is expressed in terms of a line integral along a *closed* contour that is expected to satisfy the condition, namely: [6i)]

$$\alpha_{jj+1}(\Gamma) = \oint_{\Gamma} \mathbf{ds} \cdot \boldsymbol{\tau}_{jj+1}(\mathbf{s}|\Gamma) = n_j \pi \tag{1}$$

Here $\Gamma$ is the closed contour, $\boldsymbol{\tau}_{jj+1}$ are the relevant NACTs, $\mathbf{s}$ stands for a group of nuclear coordinates, $\mathbf{ds}$ is an infinitesimal vectorial length along $\Gamma(\mathbf{s})$, the dot stands for scalar product, $\alpha_{jj+1}(\Gamma)$ is recognized as the topological phase and $n_j$ is an integer (or zero). In case the closed contour, $\Gamma$, is a *circular contour* Eq. (1) simplifies to become: [6{c,j})]

$$\alpha_{jj+1}(q) = \int_0^{2\pi} \tau_{\varphi jj+1}(\varphi|q) d\varphi \qquad (2)$$

where $q$ and $\varphi$ were mentioned earlier and $\tau_{\varphi jj+1}(\varphi|q)$ is the corresponding *angular* (tangential) component of $\tau_{jj+1}$ defined along circular arcs, thus: [6c]

$$\tau_{\varphi jk}(\varphi|q) = \left\langle \zeta_j(\varphi|q) \left| \frac{\partial}{\partial \varphi} \zeta_k(\varphi|q) \right. \right\rangle \qquad (3)$$

In addition to $\alpha_{jj+1}(q)$ we are frequently interested also in the adiabatic-to-diabatic transformation (ADT) angle $\gamma_{jj+1}(\varphi|q)$ defined along an open contour [7b]

$$\gamma_{jj+1}(\varphi|q) = \int_0^{\varphi} \tau_{\varphi jj+1}(\varphi'|q) d\varphi' \qquad (4)$$

As mentioned earlier the tri-state case is detailed in Appendix II and here are summarized the main results relevant for the present study.

To treat the tri-state case we have to consider the equation given in Eq. (II.1) which yields the ADT matrix, $\mathbf{A}(\varphi|q)$ (the matrix that replaces the ADT angles $\gamma_{jj+1}(\varphi|q)$ in the two-state case): [7b]

$$\frac{\partial}{\partial \varphi} \mathbf{A}(\varphi|q) + \boldsymbol{\tau}(\varphi|q) \mathbf{A}(\varphi|q) = \mathbf{0} \qquad (5)$$

where the anti-symmetric, 3x3, $\mathbf{\tau}(\varphi|q)$ - matrix is given in the form [42]

$$\mathbf{\tau}(\varphi|q)=\begin{pmatrix} 0 & \tau_{12} & \tau_{13} \\ -\tau_{12} & 0 & \tau_{23} \\ -\tau_{13} & -\tau_{23} & 0 \end{pmatrix} \tag{6}$$

Since the ADT matrix is known to be an orthogonal (unitary) matrix [6n] it can be expressed in terms of the following three (quasi-) Euler angles $\gamma_{12}$, $\gamma_{13}$ and $\gamma_{23}$. To be more explicit, this matrix is presented as a *product* of three rotation matrices $\mathbf{Q}_{12}(\gamma_{12})$, $\mathbf{Q}_{13}(\gamma_{13})$ and $\mathbf{Q}_{23}(\gamma_{23})$ where, for instance, $\mathbf{Q}_{13}(\gamma_{13})$ is given in the form:

$$\mathbf{Q}_{13}(\gamma_{13})=\begin{pmatrix} \cos\gamma_{13} & 0 & \sin\gamma_{13} \\ 0 & 1 & 0 \\ -\sin\gamma_{13} & 0 & \cos\gamma_{13} \end{pmatrix} \tag{7}$$

and the other two matrices are of a similar structure with the respective cosine and sine functions at the appropriate positions. The order of the three $\mathbf{Q}$-matrices in the product is, at this point, arbitrary which allows us to choose the more convenient order for each case separately.

We start with an ADT matrix which is given by the product [6{o,p},43,44]

$$\mathbf{A}(\varphi|q)=\mathbf{Q}_{12}\mathbf{Q}_{13}\mathbf{Q}_{23} \tag{8a}$$

In Appendix II we showed that substituting Eq. (8a) in Eq. (5) leads to three first-order differential equations for the respective three Euler angles.[6{o,p}] The feature that characterizes this set of equations is that it always breaks up into a set of two coupled equations, e.g., the equations for $\gamma_{12}$ and $\gamma_{13}$ (and a separate equation for $\gamma_{23}$ which is of no interest for us at this stage). Thus we encounter the two coupled equations:

$$\frac{\partial}{\partial \varphi}\gamma_{12} = -\tau_{\varphi 12} - \tan\gamma_{13}(\tau_{\varphi 23}\cos\gamma_{12} + \tau_{\varphi 13}\sin\gamma_{12})$$

$$\frac{\partial}{\partial \varphi}\gamma_{13} = \tau_{\varphi 23}\sin\gamma_{12} - \tau_{\varphi 13}\cos\gamma_{12} \tag{9a}$$

which, for a given set of boundary conditions can be solved without difficulty.

The more interesting angle in Eqs. (9a) is $\gamma_{12}(\varphi|q)$, which essentially is an *improved* value of the two-state $\gamma_{12}$: [7b]

$$\gamma_{12}(\varphi|q) = -\int_0^\varphi \tau_{\varphi 12}(\varphi'|q)d\varphi' \tag{10a}$$

(that follows from Eq. (4) assuming $j=1$). Similarly, the corresponding *improved* value of the two-state *topological phase* given by Eq. (2) for $j=1$, i.e. $\alpha_{12}(q)$ ($\equiv \gamma_{12}(\varphi=2\pi|q)$) follows from completing the integration of Eqs. (9a) up to $\varphi=2\pi$.

This approach is also applied for calculating the *improved* value of $\gamma_{23}$ (see Eq. (4) for $j=2$), namely, the situation where the two upper (excited) states of the system are disturbed by the lowest state via the NACTs: $\tau_{\varphi 12}(\varphi|q)$ and $\tau_{\varphi 13}(\varphi|q)$. In this case the **A**-matrix is assumed to be given by the product:

$$\mathbf{A} = \mathbf{Q}_{23}\mathbf{Q}_{13}\mathbf{Q}_{12} \qquad (8b)$$

and the relevant coupled differential equations for the two Euler angles $\gamma_{23}$ and $\gamma_{13}$ can be shown to be of the form (see Appendix II):

$$\frac{\partial}{\partial \varphi}\gamma_{23} = -\tau_{\varphi 23} + \tan\gamma_{13}(\tau_{\varphi 12}\cos\gamma_{23} - \tau_{\varphi 13}\sin\gamma_{23})$$

$$\frac{\partial}{\partial \varphi}\gamma_{13} = -\tau_{\varphi 12}\sin\gamma_{23} - \tau_{\varphi 13}\cos\gamma_{23} \qquad (9b)$$

Here, like in the previous case, the more interesting ADT angle is $\gamma_{23}(\varphi|q)$ which essentially is the improved value of the two-state result given by the expression (see Eq. (4) for $j=2$):

$$\gamma_{23}(\varphi|q) = -\int_0^\varphi \tau_{\varphi 23}(\varphi'|q)d\varphi' \qquad (10b)$$

Similarly, the corresponding value of the topological phase $\alpha_{23}(q)$ ($\equiv \gamma_{23}(\varphi=2\pi|q)$), which follows by completing the integration of Eqs. (9b) up to $\varphi=2\pi$, is an improved value of the two-state topological phase as calculated from Eq. (2) for $j=2$.

In Appendix II we also give another way of solving Eq. (5), namely, by presenting the matrix $\mathbf{A}(\varphi|q)$ in terms of an exponentiated line-integral along a circular contour [61]

$$\mathbf{A}(\varphi|q) = \wp \exp\left[\int_0^\varphi d\varphi\, \boldsymbol{\tau}_\varphi(\varphi|q)\right] \tag{11a}$$

Here the symbol $\wp$ is introduced to indicate that the exponentiated integration has to be carried out in a given order.

Next we introduce the relevant expression for the Topological matrix, $\mathbf{D}(q)$:

$$\mathbf{D}(q) = \wp \exp\left(-\int_0^{2\pi} \boldsymbol{\tau}_\varphi(\varphi|q) d\varphi\right) \tag{11.b}$$

where the integration is carried out along a *closed* circular contour.

## III. Numerical Study

As already mentioned earlier we intend to study three tri-atomic systems, namely $H_3$, $H_3^+$ and $H_3^{++}$ although our main efforts will be devoted to $H_3^{++}$ which is studied here for the first time. In contrast, the two other systems, $H_3^+$, $H_3$, were frequently treated on various occasions and are considered here only briefly with the aim of studying the hierarchy of the topology features that develop while moving from the doubled charged $H_3^{++}$, through the singled charged $H_3^+$ and ending with the neutral $H_3$ system.

In all three cases the calculations of the NACTs along chosen circular contours are carried out at the state-average CASSCF level employing STO-6G basis set. We used the active space with one electron, two electrons and three electrons for $H_3^{++}$, $H_3^+$ and $H_3$ respectively distributed on three orbitals. Accordingly the following three lowest states, namely, $1\,^2A'$, $2\,^2A'$, and $3\,^2A'$ for $H_3^{++}$, the three lowest states $1\,^1A'$, $2\,^1A'$, and $3\,^1A'$ for $H_3^+$ and the three lowest states $1\,^2A'$, $2\,^2A'$, and $3\,^2A'$ for $H_3$ were computed by the state-average CASSCF method with equal weights. The actual calculations were carried out employing the MOLPRO program.[45]

The above mentioned molecular systems possess numerous *ci*s. Therefore, in order to conduct a meaningful study, we limit our attention to a certain group of *ci*s which can be identified in each of the three molecular systems. Moreover this group has to be characterized by the feature that each *ci* can be found at any distance in the range $\delta r = \{1-20\}$ a.u. It turns out that the equilateral $D_{3h}$ *ci*s form such a group for the three molecular systems to be studied.

The numerical results are summarized in six Tables: In Tables I-IV are presented results for $H_3^{++}$, in Table V for $H_3$ and in Table VI for $H_3^+$. Five out of the six tables contain the same kind of results, namely, the Adiabatic potential energy *curves* (APECs) calculated as a function of $R$ (see Fig.1), the angular NACTs, $\tau_{\varphi ij}(\varphi|q)$; $i\neq j=1,2,3$ as a function of $\varphi$ for a given origin (center) and a radius $q$ (see Eq. (3)); the corresponding ADT angles, $\gamma_{12}(\varphi|q)$ and $\gamma_{23}(\varphi|q)$ (see, Eq. (9a) and (9b), respectively); the topological phases, $\alpha_{12}(q)$ and $\alpha_{23}(q)$ (in print) expected to be quantized, namely, being $n\pi$ where n is an integer (or zero); the three *diagonal* elements of the ADT-matrix, $\mathbf{A}_{jj}(\varphi|q)$, $j=\{1-3\}$ (see Eq. (11a)) as a function of $\varphi$ and the corresponding (printed) diagonal elements of the topological $\mathbf{D}$-matrix (see Eq. (11b)), $\mathbf{D}_{jj}(q)$, $j=\{1-3\}$ expected to be quantized as well, namely, being equal to $\pm 1$.

At this stage is defined the concept of a *situation*: A *situation* is a member of a *group* of configurations related to a fixed inter-atomic distance, $r$, and a given equilateral *ci*. The value of $R$ at the *ci*-point, to be designated as, $R_{D3h}$, is by definition: $R_{D3h} = r\sqrt{(0.75)} \equiv 0.866r$ (see Fig 1). All angular results are derived along circles with their center at the $D_{3h}$ ci and radius: q=0.5a.u.

## III.1 Study of the $H_3^{++}$ Molecular System

In Tables I and II are summarized the main results for $H_3^{++}$ as collected for six different situations: In Table I we present results for small and intermediate values of r, whereas in Table II are presented results for large values of r to be recognized as the *asymptotic region*. Whereas the NACTs in Table I are seen to be smooth and friendly the NACTs in Table II are spiky. The more characteristic feature encountered here is that within the considered circular region it is seen that, the NACTs form one single $ci$ – a (2,3) $ci$ – which is produced by the two upper (excited) states.

One of the issues to be discussed in this article is the quantization. The fulfillment of the quantization for a given situation is an indication that the NACTs just mentioned are mathematically well converged and therefore can be trusted as reliable magnitudes for the physical *message* of interest. Indeed tests to this effect are presented in Tables III and IV

Table III contains results related to two situations – one presented in Table I and the other in Table II – for which the calculations were repeated with significantly larger radii and significantly shifted centers away from the $D_{3h}$ $ci$-point. The first situation is treated for $r = 8.0$ a.u., $q = 1.5$ a.u. and a center located at $R_{C2v} = 8.0$ a.u. (thus shifted away by 1.07 a.u. from the $D_{3h}$ $ci$). The second situation is treated for $r = 10.0$ a.u., $q = 3.0$ a.u. and a center located at the $R_{C2v} = 11.0$ a.u. (thus shifted away by 2.34 a.u). As is noticed, the two studies presented in Table III yield, for significantly different situations as presented in Table I and II, similar *quantizations* due to the equilateral $D_{3h}$ $ci$s at $r = 8.0$ a.u. and 10.0 a.u., respectively. In other words the results presented in Table III support the corresponding results in Tables I and II that indeed the encountered NACTs form, in the corresponding configuration space (CS), one single $ci$.

The *quality* of the *quantization* for the results presented so far is further examined in Table IV. Here are compared results as derived from tri-state calculations with those derived by two-state calculations (expected to be less accurate) with the aim of establishing the convergence of the tri-state calculations. The comparison is done for both ADT angles $\gamma(\varphi|q)$ and the corresponding *topological* phases, α(*q*). Indeed the quantization in both cases is similar but those due to the tri-state calculation are more acurate.

## III.2 Study of the $H_3$ and $H_3^+$ Molecular Systems

In Tables V and VI are presented results for situations due to $D_{3h}$ *ci*s for $H_3$ along the *r*-interval {6.0, 13.0} a.u. and in case of $H_3^+$, along the *r*-interval {6.0, 15.0} a.u. In each Table we consider four situations: three of them apply for intermediate *r*-values and the fourth applies for an asymptotic case. The main reason for presenting them is to find out to what extent their intermediate topological features differ from the corresponding asymptotic ones and to what extent they differ from those just exposed for $H_3^{++}$. As for the comparison between $H_3$ and $H_3^+$ two features are noticed: (i) In both cases the $D_{3h}$ NACTs are only slightly affected – while *r* (and *R*) varies from the internal region towards the asymptotic one. (ii) Still the two molecular systems behave differently, during this process: in case of $H_3$ the encountered NACTs are associated with the (1,2) *ci* [25-27] whereas in case of $H_3^+$ they are associated with the (2,3) *ci*.

## IV. Analysis and Conclusions

The foremost purpose of the numerical treatment is to reveal to what extent the $H_3^{++}$ molecular system differs from the $H_3$ and $H_3^+$ systems, mainly topologically but eventually also in other ways. The means to carry out this study are the NACTs calculated along closed circular contours formed by the various equilateral *ci*s.

From the results presented in Tables V and VI the two systems, $H_3$ and $H_3^+$, respectively, are behaving smoothly all the way from the internal regions of the CS up to the asymptotic ones. Mathematically this implies that the relevant Schrödinger-BOH equations for these systems can be solved along the whole CS for any foresee/anticipative set of asymptotic conditions.

The $H_3^{++}$ does not seem to behave like that at all. While watching the NACTs produced at CSs surrounding different $D_{3h}$ points the following is revealed: (i) The $D_{3h}$ NACTs are significantly affected while moving from the internal region towards the asymptotic one (cf. results in Table I and Table II). (ii) Once the asymptotic region is reached we find, that the smoothly behaving NACTs formed in the internal regions are replaced by spiky NACTs to the level of becoming Dirac-$\delta$ functions.

Thus, summarizing the findings regarding the $H_3^{++}$ molecular system, the following can be said: Although along the intermediate regions of CS the behavior of this system as reflected via the contour-dependent NACTs is bearable, we encounter essentially a non-molecular conduct at its asymptotic region

In the present article three molecular systems are studied– two of them, $H_3$ and $H_3^+$ – are characteristic prototypes for chemical systems and thus both are capable of forming di-atom molecules following scattering exchange processes or also, as in case of $H_3^+$, [36] form either, $H_2$ or $H_2^+$, via dissociation. In contrast, the third molecular system, $H_3^{++}$ behaves differently: following the present study it looks like that $H_3^{++}$, is not capable of reaching its asymptotic region and therefore once created will not be able to dissociate and form its (just) two constituents. namely ($H_2^+$,$H^+$). The only way it may dissociate is by absorbing an electron and forming NACTs typical for an ordinary molecule via the reaction $H_3^{++} + e \rightarrow H_3^+$ that now may dissociate properly, in the form:

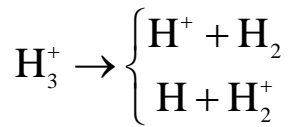

$$H_3^+ \rightarrow \begin{cases} H^+ + H_2 \\ H + H_2^+ \end{cases}$$

thus yielding ordinary di-atoms or di-atomic ions.

At this stage we return to the Introduction in which we mentioned that according to Hawking molecules seemed to appear *too early*. In the present chapter we try to use the findings revealed earlier in order examine to what extent this possibility can be justified, at least topologically.

The present study revealed the tendency of this molecule to form spiky NACTs to the level of Dirac-$\delta$ functions. Mathematically this implies that the molecular system at stake, most likely, loses its electron and breaks-up into three protons *before* reaching, chemically, its asymptotic region. If indeed that happens, the *chemical* dissociative process is inhibited and such a molecule, is not capable to behave chemically.

This study was performed with the aim of showing that, eventually, tri-atomic molecules could be formed during at or shortly after the Big Bang. In the present study we referred to the more primitive tri-atomic molecular system, $H_3^{++}$, for which we showed that although it satisfies, up to a certain level, topological features as required by the BOH theory it cannot be considered as a normal molecule because its chemical asymptotic region is essentially out of reach. Still its two unique features, namely, being formed by three protons and *one* electron and behave as a molecule - topologically - on borrowed time makes it a candidate for becoming the *cornerstone* to form molecules.

# V.   Corollary – NACTs as Gluons

In order to discuss the buildup of protons Gell-mann and Zweig suggested that Hadrons (e.g. protons) are made out of smaller particles called Quarks. These quarks (usually three of them) are assumed to be hold together by particles – Gluons – that convey the force between quarks. Such systems are discussed within the Quantum-Chromo-Dynamics approach.

In the present talk we face a similar problem, viz., building molecular systems out of Protons and Electrons. However here the BOH approach supplies us with the means to build the required magnitudes– the  NACTs – as discussed recently [46]

Thus the NACTs convey the force between nuclei that form the molecule.

# Acknowledgements

Bijit Mukherjee (File No: 09/080(0960)/2014-EMR-I) acknowledges CSIR India for research fellowship. Satrajit Adhikari acknowledges DST, India, through project no. File No. EMR/2015/001314 for the research funding. Satrajit Adhikari also acknowledges IACS for CRAY supercomputing facility.

# Appendix I

# Background Information

In order to get an idea how NACTs are expected to affect the dynamics of the nuclei within molecular systems we briefly present, for the sake of completeness, the relevant Schrödinger equation that governs the motion of the nuclei following the elimination of the electrons.

The starting point is the complete Hamiltonian of both the nuclei and the electrons: [6d]

$$\mathbf{H}(\mathbf{s_e},\mathbf{s}) = \mathbf{T_n}(\mathbf{s}) + \mathbf{H_e}(\mathbf{s_e}|\mathbf{s}) \tag{I.1}$$

where $\mathbf{H_e}(\mathbf{s_e}|\mathbf{s})$ is the electronic Hamiltonian which also contains the nuclear Coulomb interactions, $\mathbf{s_e}$ and $\mathbf{s}$ stand for sets of the electronic and nuclear coordinates, respectively, and $\mathbf{T_n}(\mathbf{s})$ is the nuclear kinetic energy written (in terms of mass-scaled coordinates) as:

$$\mathbf{T_n} = -\frac{\hbar^2}{2M}\nabla^2 \tag{I.2}$$

Here M is the mass of the system and $\nabla$ is the gradient (vector) operator expressed in terms of *mass-scaled* nuclear coordinates)

The Schrödinger equation to be considered is of the form:

$$\left(\mathbf{H}(\mathbf{s_e},\mathbf{s}) - E\right)\left|\mathbf{\Psi}(\mathbf{s_e},\mathbf{s})\right\rangle = 0 \tag{I.3}$$

where $E$ is the total energy of the system and $|\Psi(\mathbf{s}_e,\mathbf{s})\rangle$ is the complete wave function which describes the motion of both the electrons and the nuclei. Next we introduce the Born-Oppenheimer expansion: [6d]

$$|\Psi(\mathbf{s}_e,\mathbf{s})\rangle = \sum_{j=1}^{N} |\zeta_j(\mathbf{s}_e|\mathbf{s})\rangle \chi_j(\mathbf{s}) \qquad (\text{I.4})$$

where N is the dimension of the corresponding Hilbert space and $\chi_j(\mathbf{s})$, $j = \{1,N\}$ are the nuclear-coordinate dependent coefficients (recognized later as the nuclear wave functions). Next are mentioned the related electronic (adiabatic) eigen-functions $|\zeta_j(\mathbf{s}_e|\mathbf{s})\rangle$; $j = \{1,N\}$ of $\mathbf{H}_e(\mathbf{s}_e|\mathbf{s})$ given in Eq. (I.1) defined through the Eigen-value problem:

$$\left(\mathbf{H}_e(\mathbf{s}_e|\mathbf{s}) - u_j(\mathbf{s})\right)|\zeta_j(\mathbf{s}_e|\mathbf{s})\rangle = 0; \quad j=\{1,N\} \qquad (\text{I.5})$$

where $u_j(\mathbf{s})$, $j=\{1,N\}$ are the corresponding electronic eigen-values which are recognized as the adiabatic PESs governing the motion of the *nuclei*.

Substituting Eq. (I.4) in Eq. (I.3), while recalling Eq. (I.1), (I.2) and (I.5), multiplying the outcome from the left by $\langle \zeta_k(\mathbf{s}_e|\mathbf{s})|$ and integrating over the electronic coordinates yield the following set of coupled equations:

$$-\frac{\hbar^2}{2M}\nabla^2 \chi_k + (u_k - E)\chi_k - \frac{\hbar^2}{2M}\sum_{j=1}^{N}\left(2\boldsymbol{\tau}_{kj}\cdot\nabla + \boldsymbol{\tau}_{kj}^{(2)}\right)\chi_j = 0; \quad k=\{1,N\} \qquad (\text{I.6})$$

where $\boldsymbol{\tau}$ is the (first order) non-adiabatic coupling (vector) matrix with the elements:

$$\boldsymbol{\tau}_{jk} = \langle \zeta_j | \nabla \zeta_k \rangle \tag{I.7a}$$

and $\boldsymbol{\tau}^{(2)}$ is the non-adiabatic (scalar) matrix of the second order with the elements:

$$\boldsymbol{\tau}_{jk}^{(2)} = \langle \zeta_j | \nabla^2 \zeta_k \rangle \tag{I.7b}$$

For a system of real electronic wave functions $\boldsymbol{\tau}$ is an anti-symmetric matrix.

Eq. (I.6) can also be written in a matrix form as follows:

$$-\frac{\hbar^2}{2M}\nabla^2 \Theta + (\mathbf{u} - E)\Theta - \frac{\hbar^2}{2M}\left(2\boldsymbol{\tau}\cdot\nabla + \boldsymbol{\tau}^{(2)}\right)\Theta = \mathbf{0} \tag{I.8}$$

Or, more compactly:

$$-\frac{\hbar^2}{2M}(\nabla + \boldsymbol{\tau})^2 \Theta + (\mathbf{u} - E)\Theta = \mathbf{0} \tag{I.9}$$

where $\boldsymbol{\Theta}(\mathbf{s})$ is a column-vector that contains the above mentioned nuclear wave functions $\chi_j(\mathbf{s})$, $j=\{1,N\}$ and $\mathbf{u}$ is a *diagonal* matrix which contains the corresponding adiabatic PESs, $u_j(\mathbf{s})$, introduced via Eq. (I.5).

Next we consider the Adiabatic-to-Diabatic Transformation (ADT) [6i] which is fulfilled by replacing $\boldsymbol{\Theta}(\mathbf{s})$ in Eq. (I.9) by another column vector $\boldsymbol{\Phi}(\mathbf{s})$:

$$\Theta(\mathbf{s}) = \mathbf{A}(\mathbf{s})\Phi(\mathbf{s}) \qquad (\text{I}.10)$$

where $\mathbf{A}(\mathbf{s})$ is the ADT matrix to be determined by the requirement that the $\tau$-matrix in Eq. (I.9) is eliminated. This happens if $\mathbf{A}(\mathbf{s})$ fulfills the following first order differential equation: [6k]

$$\nabla \mathbf{A}(\mathbf{s}) + \tau(\mathbf{s})\mathbf{A}(\mathbf{s}) = \mathbf{0} \qquad (\text{I}.11)$$

which causes Eq. (I.10) to become:

$$-\frac{\hbar^2}{2M}\nabla^2 \Phi + (\mathbf{W} - E)\Phi = \mathbf{0} \qquad (\text{I}.12)$$

where $\mathbf{W}$ is the corresponding diabatic potential matrix given in the form:

$$\mathbf{W}(\mathbf{s}) = \mathbf{A}(\mathbf{s})^\dagger \mathbf{u}(\mathbf{s})\mathbf{A}(\mathbf{s}) \qquad (\text{I}.13)$$

# Appendix II

# Treatment of the Tri-State Case: Introducing the Euler Angles

The starting point is Eq. (I.11) which yields the Adiabatic-to-Diabatic Transformation (ADT) [6k] matrix $\mathbf{A(s)}$. For a circular (planar) contour this equation takes the form:

$$\frac{\partial}{\partial \varphi}\mathbf{A}(\varphi|q) + \boldsymbol{\tau}(\varphi|q)\mathbf{A}(\varphi|q) = \mathbf{0} \tag{II.1}$$

where $\boldsymbol{\tau}_\varphi(\varphi|q)$ is the matrix which contains the individual NACTs defined as: [6c]

$$\boldsymbol{\tau}_{\varphi jk}(\varphi|q) = \left\langle \zeta_j(\varphi|q) \left| \frac{\partial}{\partial \varphi} \zeta_k(\varphi|q) \right. \right\rangle \tag{II.2}$$

The solution of Eq. (II.1) is usually carried out along a contour $\Gamma$ and can be presented in terms of an exponentiated line-integral: [6l]

$$\mathbf{A}(\varphi|q) = \wp \exp\left[ \int_0^\varphi d\varphi \boldsymbol{\tau}_\varphi(\varphi|q) \right] \tag{II.3}$$

Here the symbol $\wp$ is introduced to indicate that the exponentiated integration has to be carried out in a given order.

The relevant Topological matrix $\mathbf{D}(q)$ is derived by completing the integral in Eq. (II.3) for a closed contour. [6l] Thus:

$$\mathbf{D}(q) = \wp \exp\left( -\int_0^{2\pi} \boldsymbol{\tau}_\varphi(\varphi|q) d\varphi \right) \tag{II.4}$$

The feature most associated with **D**-matrix is the *quantization* which was mentioned earlier while discussing the two-state case [6j] (Eq. (1) – see main text). In the tri-state case the quantization is satisfied when **D**($\Gamma$) becomes *diagonal* with (+1)s and (−1)s along its diagonal for any contour in the assigned region in configuration space. [6m] In what follows the number of (−1)s is labeled as K, and is defined as the topological number.[6m] The importance of K stems from the fact that it yields the number of eigenfunctions that flip sign while the electronic manifold traces a closed contour $\Gamma$. This makes the number K, contour-dependent i.e. K=K($\Gamma$) with one limitation, namely, K has to be an even number (or zero) implying that at each such a calculation an even number of electronic eigen-functions (in the Jahn-Teller sense) flip their sign. The matrix **D**($\Gamma$) is not only characterized by the number K but also by the positions of the (−1)s along its diagonal. [6m] It was shown that the position of a given (−1) corresponds to a particular eigen-function that flips its sign when the electronic manifold traces the contour $\Gamma$.

To continue we consider the tri-state system. Here the 3x3 matrix-$\boldsymbol{\tau}(\mathbf{s})$ can be shown to be of the form:

$$\boldsymbol{\tau}(\mathbf{s}) = \begin{pmatrix} 0 & \tau_{12} & \tau_{13} \\ -\tau_{12} & 0 & \tau_{23} \\ -\tau_{13} & -\tau_{23} & 0 \end{pmatrix} \quad \text{(II.7)}$$

For this case we consider the corresponding ADT matrix **A**(**s**) which is known to be an orthogonal (unitary) matrix [6n] and therefore can be expressed in terms of the following three quasi-Euler angles $\gamma_{12}$, $\gamma_{13}$ and $\gamma_{23}$. To be more explicit, this matrix is presented as a product of three rotation matrices $\mathbf{Q}_{12}(\gamma_{12})$, $\mathbf{Q}_{13}(\gamma_{13})$ and $\mathbf{Q}_{23}(\gamma_{23})$ where, for instance, $\mathbf{Q}_{13}(\gamma_{13})$ is given in the form:

$$\mathbf{Q}_{13}(\gamma_{13}) = \begin{pmatrix} \cos\gamma_{13} & 0 & \sin\gamma_{13} \\ 0 & 1 & 0 \\ -\sin\gamma_{13} & 0 & \cos\gamma_{13} \end{pmatrix} \tag{II.8}$$

and the other two matrices are of a similar structure with the respective cosine and sine functions at the appropriate positions. The order of the three **Q**-matrices in the product is, at this point, arbitrary which allows us to choose the more order for our purposes

We start with an ADT matrix which follows from the product [60,p]

$$\mathbf{A} = \mathbf{Q}_{12}\mathbf{Q}_{13}\mathbf{Q}_{23} \tag{II.8a}$$

and therefore takes the explicit form:

$$\mathbf{A} = \begin{pmatrix} c_{12}c_{13} & s_{12}c_{23} - c_{12}s_{13}s_{23} & s_{12}s_{23} + c_{12}s_{13}c_{23} \\ -s_{12}c_{13} & c_{12}c_{23} + s_{12}s_{13}s_{23} & c_{12}s_{23} - s_{12}s_{13}c_{23} \\ -s_{13} & -c_{13}s_{23} & c_{13}c_{23} \end{pmatrix} \tag{II.9}$$

where $c_{kj} = \cos(\gamma_{kj})$ and $s_{kj} = \sin(\gamma_{kj})$.

Substituting Eq. (II.9) in Eq. (II.1) leads to three first-order differential equations with respect to the three Euler angles.[60,p] The feature that characterizes these three coupled equations is the fact that they always break-up into a set of two coupled equations e.g., the equations for $\gamma_{12}$ and $\gamma_{13}$ (and one separate equation for $\gamma_{23}$ which is of no interest for us in the present study). The two coupled equations are: [60]

$$\frac{\partial}{\partial \varphi}\gamma_{12} = -\tau_{\varphi 12} - \tan\gamma_{13}(\tau_{\varphi 23}\cos\gamma_{12} + \tau_{\varphi 13}\sin\gamma_{12})$$

$$\frac{\partial}{\partial \varphi}\gamma_{13} = \tau_{\varphi 23}\sin\gamma_{12} - \tau_{\varphi 13}\cos\gamma_{12} \tag{II.11a}$$

which, for a given set of boundary conditions, can be solved without difficulty.

This approach can be applied also for calculating the extended value of $\gamma_{23}$ (see Eq. (4) for $j=2$, in the text) namely, the case where the two upper (excited) states of the system are not isolated but are disturbed by the lowest state via the NACTs: $\tau_{\varphi 12}(\varphi|q)$ and $\tau_{\varphi 13}(\varphi|q)$. In this case the **A**-matrix is written as:

$$\mathbf{A} = \mathbf{Q}_{23}\mathbf{Q}_{13}\mathbf{Q}_{12} \tag{II.8b}$$

and the relevant coupled differential equations for the two Euler angles, $\gamma_{23}$ and $\gamma_{13}$ are given in the form:

$$\frac{\partial}{\partial \varphi}\gamma_{23} = -\tau_{\varphi 23} + \tan\gamma_{13}(\tau_{\varphi 12}\cos\gamma_{23} - \tau_{\varphi 13}\sin\gamma_{23})$$

$$\frac{\partial}{\partial \varphi}\gamma_{13} = -\tau_{\varphi 12}\sin\gamma_{23} - \tau_{\varphi 13}\cos\gamma_{23} \tag{II.11b}$$

In previous publications [6p] the two sets of coupled equations, namely Eqs. (II.11a) and (II.11b), were termed as The Coupled Equations with Privilege: Eq. (II.11a) is privileged with the (1,2) NACT, $\tau_{\varphi 12}(\varphi|q)$ and Eqs. (II.11b) is privileged with the (2,3) NACT, $\tau_{\varphi 23}(\varphi|q)$.

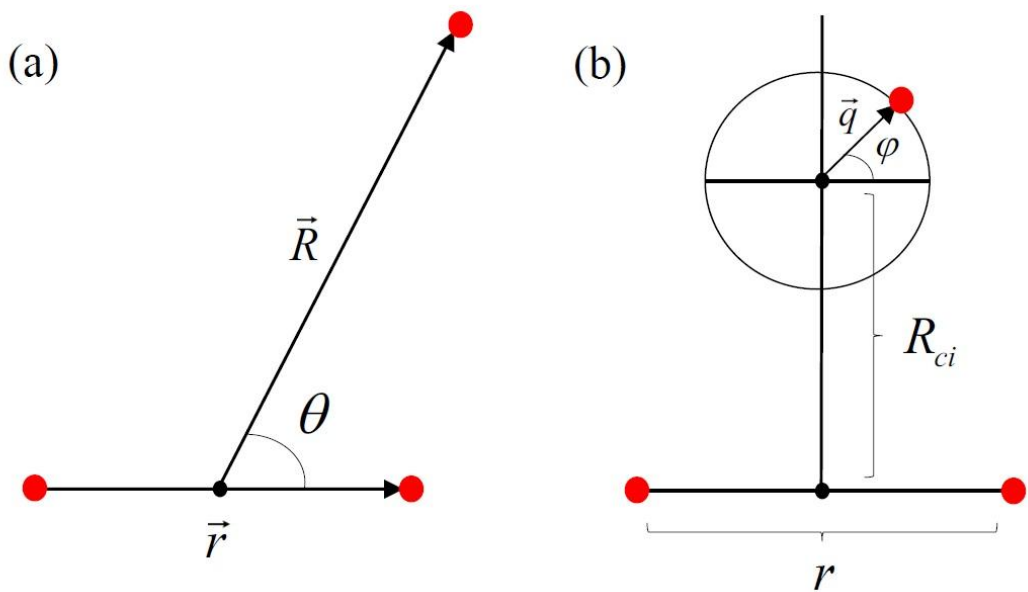

**Figure 1.** Schematic picture of the system of coordinates and various points of locations: (a) The Jacobi system of coordinates, namely, $r$, $R$ and $\theta$; (b) the point of (1,2) or (2,3) $ci$ is at $R = R_{ci}$ and $\theta = \pi/2$, which is the center of the circle with the radius $q$. The coordinates $(R_{ci}, \theta \mid r)$ and $(\varphi, q \mid r)$ are used throughout the study. The three hydrogen nuclei are represented by red circles.

# Table I

## Results for $H_3^{++}$ Molecular System
### (intermediate *r* - values)

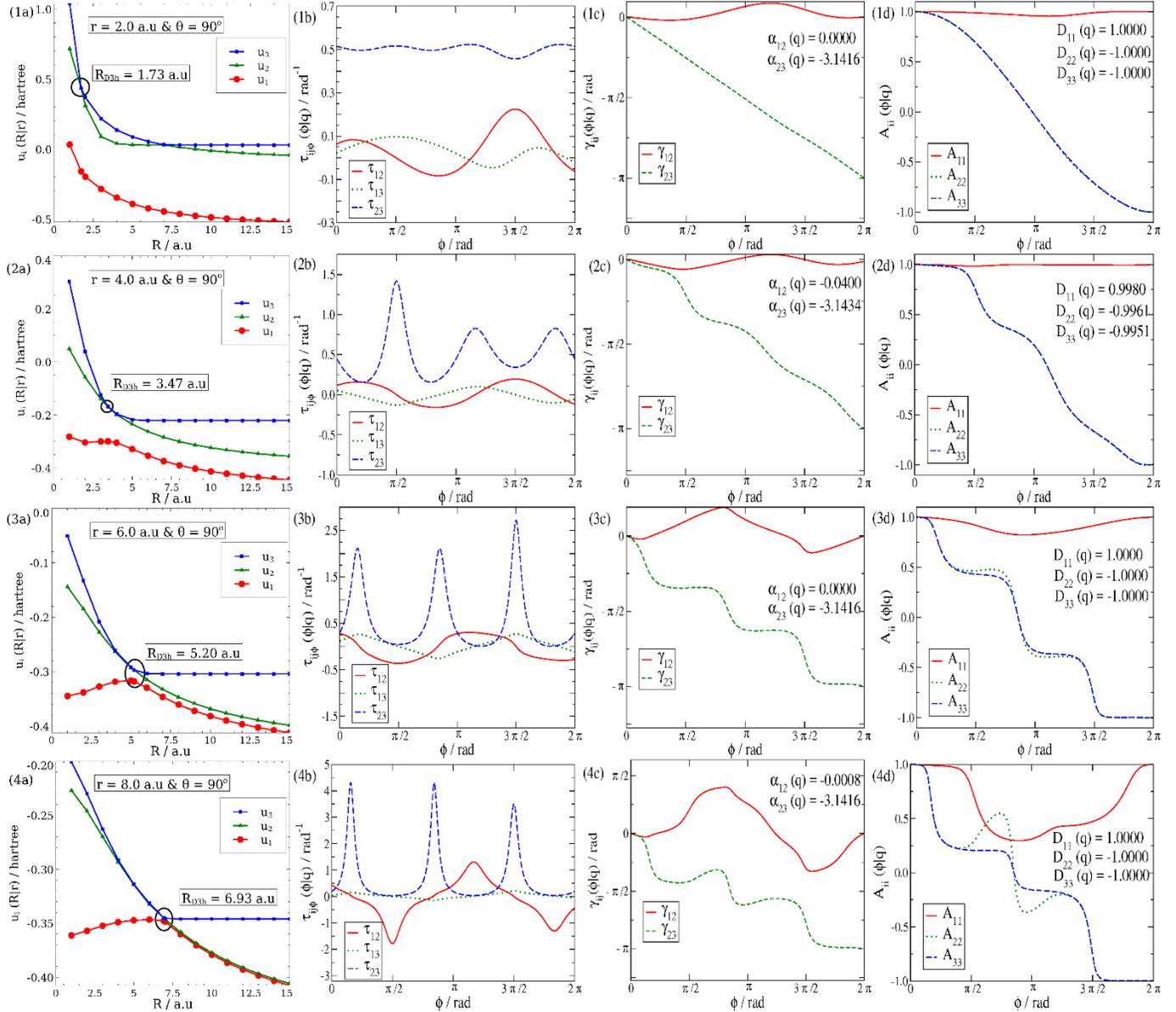

**Table I:** Results in each row are calculated for a given *r*-value: $r = 2.0, 4.0, 6.0$ and $8.0$ a.u. In sub-figures (a) are presented three APECs, $u_i(R|r)$, $i=\{1,3\}$ related to the three lowest eigenstates as a function of $R$. Shown are also the positions of the relevant equilateral $D_{3h}$ (2,3) *ci* presented by $R_{D3h}$. In sub-figures (b) are presented the three relevant, NACTs, $\tau_{\varphi ij}$ $(\varphi|q)$ as a function of $\varphi$, calculated for circles located at the relevant (2,3) $D_{3h}$ *ci*s and radius $q = 0.5$ a.u. In sub-figures (c) are presented the two relevant ADT angles $\gamma_{12}(\varphi|q)$ and $\gamma_{23}(\varphi|q)$ as a function of $\varphi$ usually very nicely quantized and the relevant (printed) topological phases $\alpha_{12}(q)$ and $\alpha_{23}(q)$. In sub-figures (d) are presented the three diagonal elements of the ADT matrix $\mathbf{A}_{jj}$ $(\varphi|q)$; $j=\{1,3\}$ as a function of $\varphi$ and the (printed) relevant topological **D**-matrix elements. $\mathbf{D}_{jj}(q)$; $j=\{1,3\}$.

# Table II

## Results for $H_3^{++}$ Molecular System
### (coalesces situations in the asymptotic region)

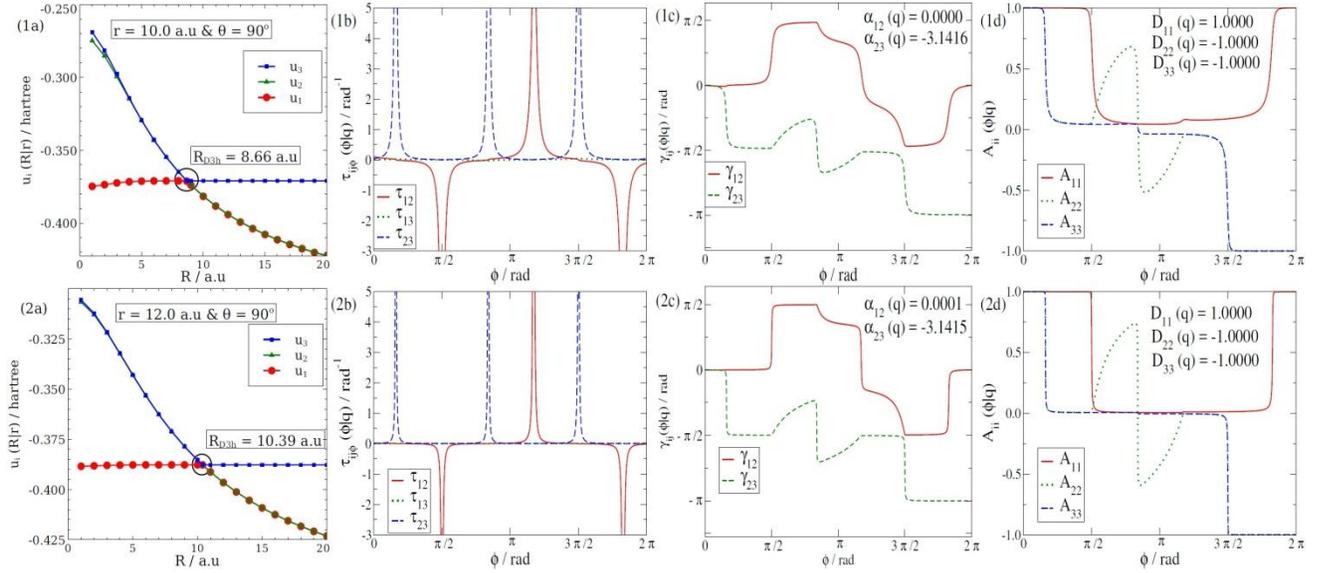

**Table II:** Results in each row are calculated for the $r$-values: $r = 10.0$ and $12.0$ a.u. The rest is as in Table I.

# Table III

## Results for $H_3^{++}$ Molecular System
### (calculated along different circular contours)

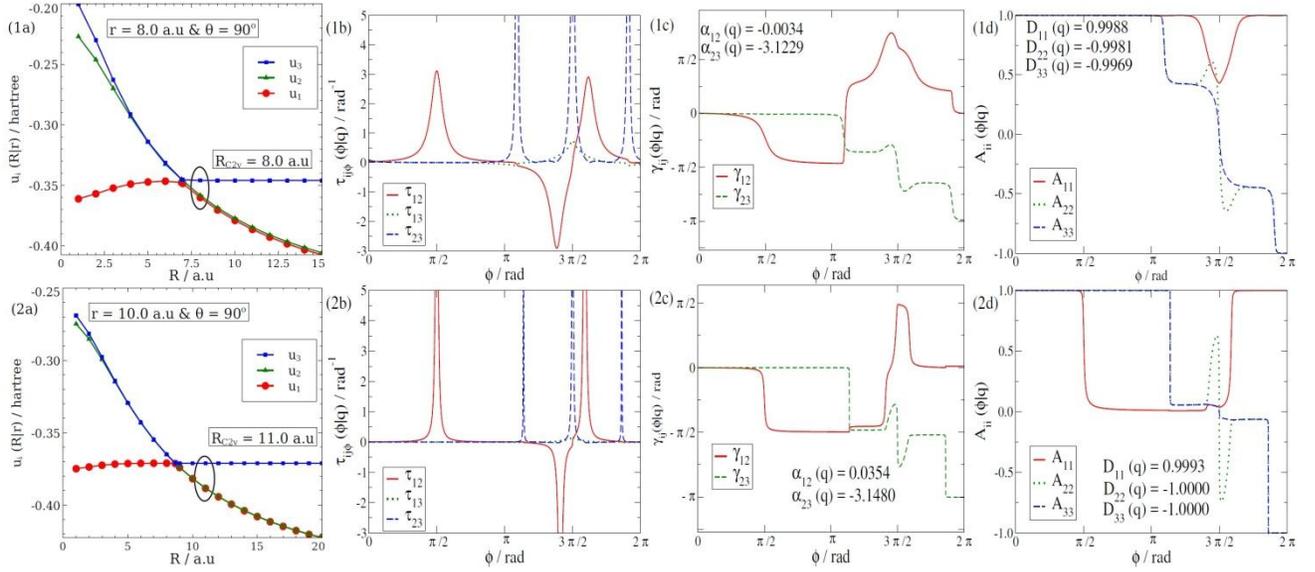

**Table III:** Results in the first row are calculated for $r = 8.0$ a.u., $q = 1.5$ a.u. and *removed* from the equilateral $D_{3h}$ (2,3) *ci* by $\Delta R = 1.07$ a.u. Results in the second row are calculated for $r = 10.0$ a.u., $q = 3.0$ a.u. and *removed* from the equilateral $D_{3h}$ (2,3) *ci* by $\Delta R = 2.34$ a.u. The rest is as in Table I.

# Table IV

## Topological Convergence Tests for $H_3^{++}$ Molecular System

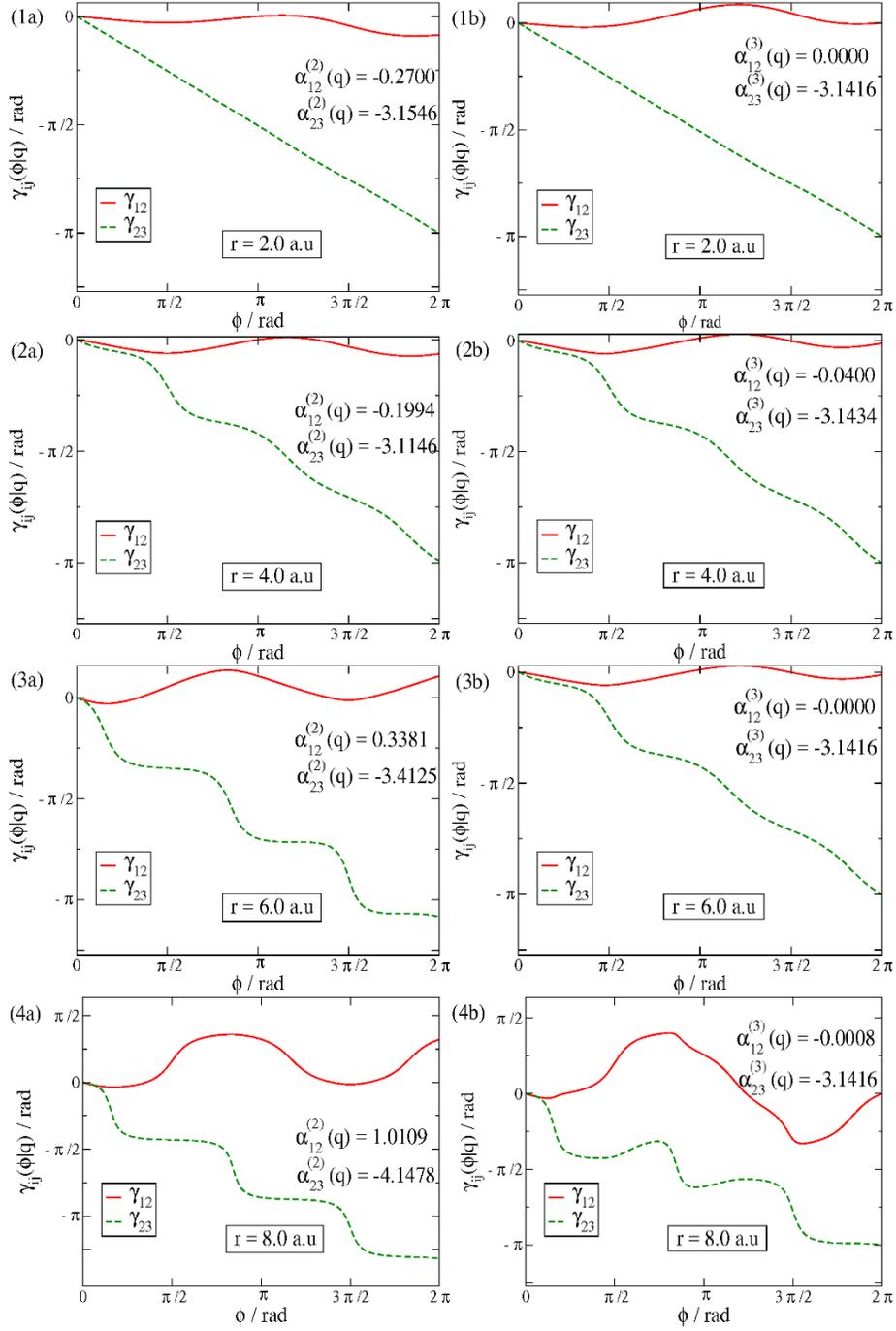

**Table IV:** Convergence tests for the ADT angles, $\gamma_{12}(\varphi|q)$ and $\gamma_{23}(\varphi|q)$ and the topological phases, $\alpha_{12}(q)$ and $\alpha_{23}(q)$, as obtained by comparing two-state (first column) and tri-state (second column) results for four different $r$-values: $r = 2.0$, 4.0, 6.0 and 8.0 a.u. Subfigure (a) of each panel displays the ADT angles ($\gamma_{12}^{(2)}(\varphi|q)$ and $\gamma_{23}^{(2)}(\varphi|q)$) obtained from two-state ADT calculations, whereas subfigure (b) of each panel exhibits ADT angles ($\gamma_{12}^{(3)}(\varphi|q)$ and $\gamma_{23}^{(3)}(\varphi|q)$) as obtained considering tri-state calculations. Also, the corresponding topological angles, $\alpha_{ij}^{(2)}(q)$ and $\alpha_{ij}^{(3)}(q)$ are shown in subfigures (a) and (b) of each panel, respectively.

# Table V

## Results for H₃ Molecular System

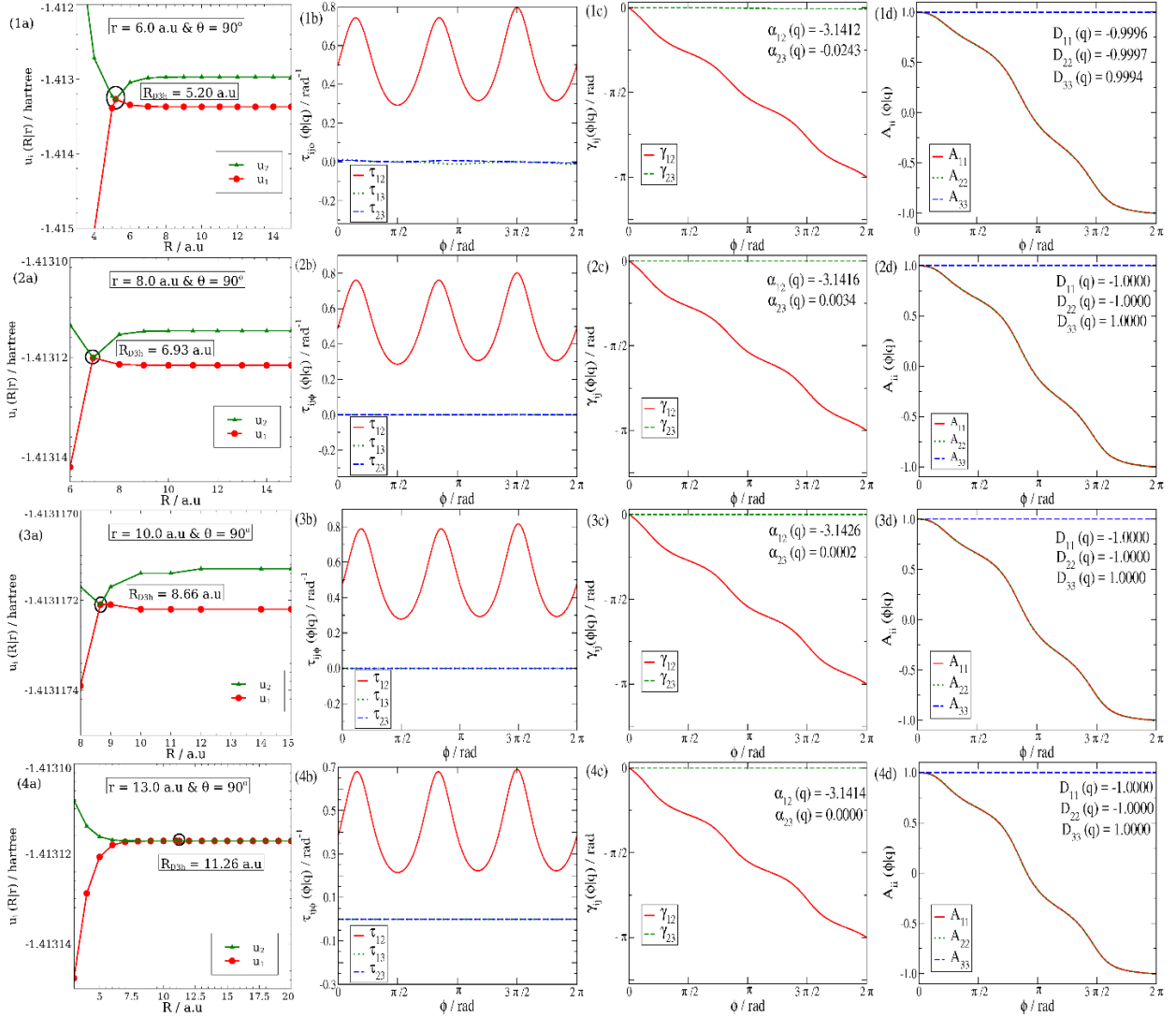

**Table V:** Results in each row are calculated for a given *r*-value: $r = 6.0, 8.0, 10.0$ and $13.0$ a.u. Since the third APES is far above the lower two APESs, it is not shown for convenience in the subfigures (a) of each panel. The rest is as in Table I.

# Table VI

## Results for $H_3^+$ Molecular System

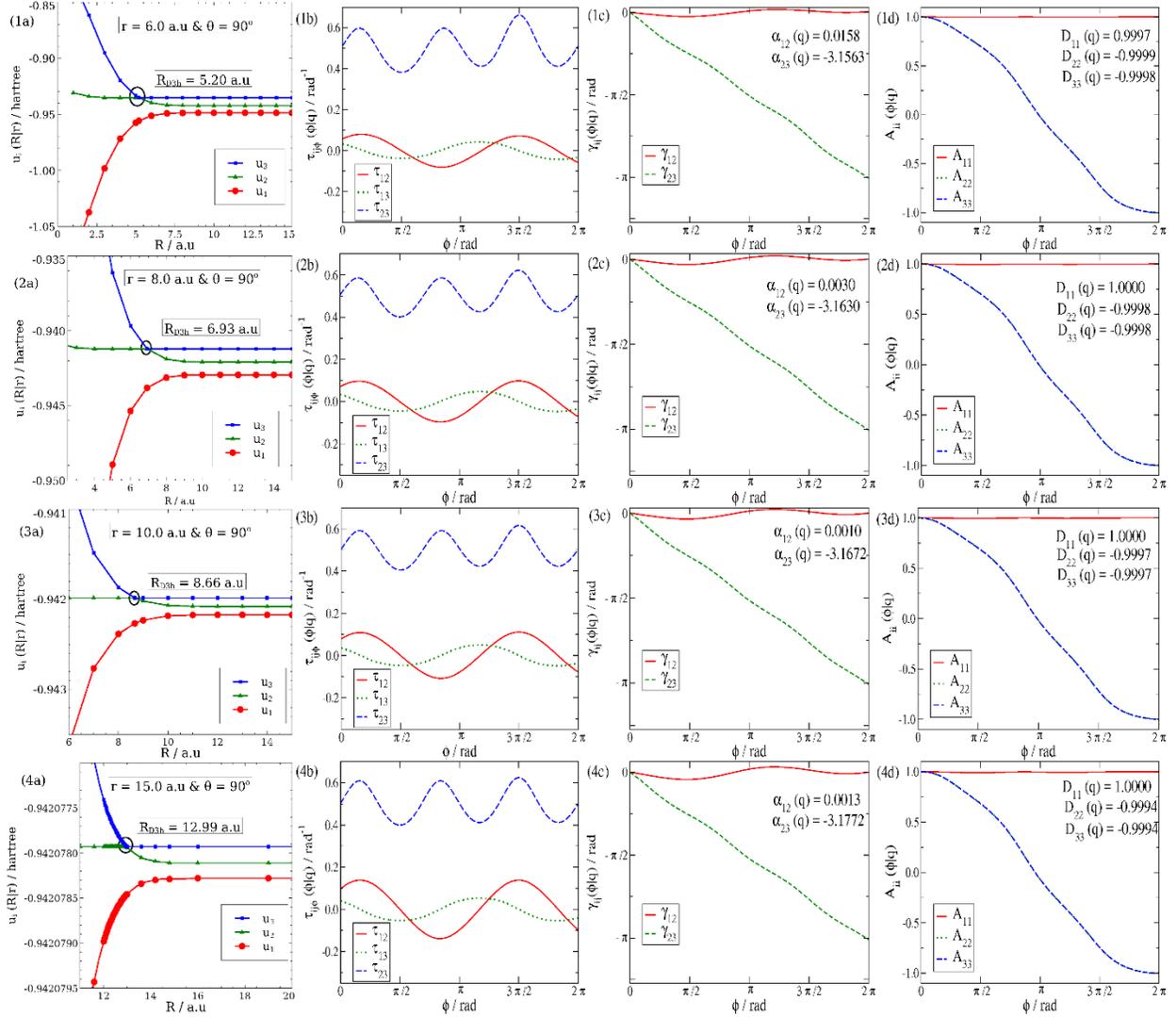

**Table VI:** Results in each row are calculated for a given *r*-value: *r* = 6.0, 8.0, 10.0 and 15.0 a.u. The rest is as in Table I.

36